\documentclass[a4paper,aps,twocolumn,nofootinbib]{revtex4}
\RequirePackage[colorlinks,hyperindex]{hyperref}
\RequirePackage[english]{babel}
\RequirePackage[latin1]{inputenc}
\RequirePackage[T1]{fontenc}
\RequirePackage{mathrsfs}
\RequirePackage{amsmath}
\RequirePackage{amssymb}
\RequirePackage{amsbsy}
\RequirePackage{color}
\RequirePackage{bm}
\hypersetup{colorlinks=true,breaklinks=true,urlcolor=blue,linkcolor=red}
\begin{document}
\title{\bf{ELKO in Polar Form}}
\author{Luca Fabbri}
\affiliation{DIME Sez. Metodi e Modelli Matematici, Universit\`{a} di Genova,\\
via all'Opera Pia 15, 16145 Genova, ITALY}
\date{\today}
\begin{abstract}
In this paper, we consider the theory of ELKO written in their polar form, in which the spinorial components are converted into products of a real module times a complex unitary phase while the covariance under spin transformations is still maintained: we derive an intriguing conclusion about the structure of ELKO in their polar decomposition when seen from the perspective of a new type of adjunction procedure defined for ELKO themselves. General comments will be given in the end.
\end{abstract}
\maketitle
\section{Introduction}
As it is well known, general spinor fields are classified in terms of the so-called Lounesto classification according to two classes: singular spinor fields are those subject to the conditions $i\overline{\psi}\boldsymbol{\pi}\psi\!=\!0$ and $\overline{\psi}\psi\!=\!0$ while regular spinor fields are all those for which the two above conditions do not identically hold \cite{L}. Broadly speaking, regular spinor fields are the Dirac spinor fields and their special cases.

Singular spinor fields can be classified in terms of further sub-classes: Weyl spinors are those that are subject to the condition $2i\overline{\psi}\boldsymbol{\sigma}_{ij}\psi\!=\!0$ with Majorana spinors being subject to $\overline{\psi}\boldsymbol{\gamma}_{a}\boldsymbol{\pi}\psi\!=\!0$ while flag-dipole spinors are all those for which the two above conditions do not identically hold \cite{Cavalcanti:2014wia,HoffdaSilva:2017waf,daSilva:2012wp,daRocha:2008we,Ablamowicz:2014rpa}. Flag-dipole spinor fields are less known in contemporary literature, although an interested reader may have a look at \cite{Vignolo:2011qt, daRocha:2013qhu} for some general treatment.

Weyl spinors and Majorana spinors are the most constrained classes. Weyl spinors are what represents massless states; Majorana spinors are what represents neutral states. But Majorana spinors have an additional feature in the fact that Dirac types of spinor field equations fail to have massive solutions in the form of plane waves.

This situation has four ways out: one is of course that Majorana spinors may not exist; a second is that if they exist they would not be massive; the third comes to rescue if we want them to be massive by requiring that they have no solution in plane waves; but if we want to save the only way we know to construct their quantum theory then we must require another type of field equation.

This last solution has been adopted by Ahluwalia and co-workers \cite{Ahluwalia:2004sz, Ahluwalia:2004ab}, who established that the simplest field equations would have to be a Klein-Gordon type of field equation. In such a case, these spinors would have a mass dimension equal to $1$ \cite{Ahluwalia:2016rwl}. They are called ELKO from the acronym in German of \emph{charge-conjugated spinors} \cite{Ahluwalia:2016jwz}.

Some studies high-light the analogies \cite{daRocha:2007pz, HoffdaSilva:2009is, Cavalcanti:2014uta} and the differences \cite{Bernardini:2005wh, Bernardini:2012sc} between Dirac spinors and ELKO.

ELKO dedicated works are for example in \cite{Villalobos:2015xca, daRocha:2011yr}.

In this paper, we will try to study ELKO in terms of the so-called polar decomposition of spinor fields. 

The polar decomposition is the process through which each of the spinorial components, as a complex number, can be written in terms of a real module times a complex unitary phase, in such a way that the covariance under a local spinor transformation is still respected \cite{Rodrigues:2005yz}.

Such a polar decomposition has been studied in \cite{Fabbri:2016msm} in a very general framework comprising regular and singular spinor fields. With a special attention for regular spinors, the polar decomposition of the Dirac spinor has then been transmitted onto the Dirac spinor field equation in \cite{Fabbri:2016laz}.

For a general review the reader may look at \cite{Fabbri:2017pwp}, while in \cite{Fabbri:2017xyk} a study of singular spinors has been done although in a rather incomplete manner, which should be settled.

It is here our aim to consider the polar decomposition of spinors done for Dirac fields in \cite{Fabbri:2018crr} in a more detailed way, whether they are spinors of Dirac or singular, whether flag-dipole or Weyl or Majorana, and ELKO.

We will witness that it seems possible for a Weyl spinor to commute into a Majorana spinor, and we will also see that neither of these spinors when written in polar form will appear to possess any degree of freedom at all.

And for ELKO we will employ the polar decomposition to study their structure, but we will find that quite surprisingly it is not possible for the ELKO to obtain a form that can polarize both helicities simultaneously.
\section{Spinor Fields in General Cases}
To set the stage, we recall that $\boldsymbol{\gamma}^{a}$ are Clifford matrices, from which $\left[\boldsymbol{\gamma}_{a}\!,\!\boldsymbol{\gamma}_{b}\right]\!=\!4\boldsymbol{\sigma}_{ab}$ and $2i\boldsymbol{\sigma}_{ab}\!=\!\varepsilon_{abcd}\boldsymbol{\pi}\boldsymbol{\sigma}^{cd}$ are the definitions of the $\boldsymbol{\sigma}_{ab}$ and the $\boldsymbol{\pi}$ matrix (which is usually indicated as a gamma with an index five, but because for the space-time this index has no meaning we shall use a notation in which there are no indices to be displayed).

It is possible to prove that these Clifford matrices verify
\begin{eqnarray}
&\boldsymbol{\gamma}_{i}\boldsymbol{\gamma}_{j}\boldsymbol{\gamma}_{k}
=\boldsymbol{\gamma}_{i}\eta_{jk}-\boldsymbol{\gamma}_{j}\eta_{ik}+\boldsymbol{\gamma}_{k}\eta_{ij}
+i\varepsilon_{ijkq}\boldsymbol{\pi}\boldsymbol{\gamma}^{q}
\end{eqnarray}
which is a spinorial matrix identity (notice that this identity shows the pseudo-scalar character of the $\boldsymbol{\pi}$ matrix).

This allows one to prove $\boldsymbol{\sigma}_{ab}$ are complex matrices verifying the Lorentz algebra, which can therefore be taken as the generators for the complex Lorentz group; because the parameters can be point-dependent, then it is a gauge group: it is known as local spin group $\boldsymbol{S}$ and the objects $\psi$ transforming as $\psi\rightarrow\boldsymbol{S}\psi$ are called spinor fields. It is possible to define an adjoint spinor field $\overline{\psi}\!=\!\psi^{\dagger}\boldsymbol{\gamma}^{0}\boldsymbol{A}$ so to have it transform as $\overline{\psi}\rightarrow\overline{\psi}\boldsymbol{S}^{-1}$ because in such a way
\begin{eqnarray}
&2\overline{\psi}\boldsymbol{\sigma}^{ab}\boldsymbol{\pi}\psi\!=\!\Sigma^{ab}\\
&2i\overline{\psi}\boldsymbol{\sigma}^{ab}\psi\!=\!M^{ab}\\
&\overline{\psi}\boldsymbol{\gamma}^{a}\boldsymbol{\pi}\psi\!=\!S^{a}\\
&\overline{\psi}\boldsymbol{\gamma}^{a}\psi\!=\!U^{a}\\
&i\overline{\psi}\boldsymbol{\pi}\psi\!=\!\Theta\\
&\overline{\psi}\psi\!=\!\Phi
\end{eqnarray}
are all objects that transform in terms of the real Lorentz group: if we also require all of them to be real-valued then we have $\boldsymbol{\gamma}^{0}\boldsymbol{A}\boldsymbol{S}^{-1}\!=\!\boldsymbol{S}^{\dagger}\boldsymbol{\gamma}^{0}\boldsymbol{A}$ or equivalently $[\boldsymbol{S},\boldsymbol{A}]\!=\!0$ for all spin transformations. Since in $4$-dimensional Clifford algebras the only matrices that commute with $\boldsymbol{S}$ and thus with all generators $\boldsymbol{\sigma}^{ab}$ are $\boldsymbol{\pi}$ and $\mathbb{I}$ then we can have that either $\boldsymbol{A}\!=\!\boldsymbol{\pi}$ or $\boldsymbol{A}\!=\!\mathbb{I}$ only. However, the two cases would simply switch each parity-even bi-linear quantity and the corresponding parity-odd bi-linear quantity according to $\Sigma^{ab}\!\leftrightarrow\!M^{ab}$ and $S^{a}\!\leftrightarrow\!U^{a}$ and $\Theta\!\leftrightarrow\!\Phi$ and this means that both cases give the same bi-linear quantities up to a mere re-definition. Thus we simply pick $\overline{\psi}\!=\!\psi^{\dagger}\boldsymbol{\gamma}^{0}$ to define the adjoint. Nevertheless, we remark that generalizations in the definition of adjoint, and so of the bi-linear quantities, are possible in different contexts \cite{Crawford:1990qf,Crawford:1991mm}. However, in this paper we will focus only on the definitions given above.

For these bi-linear quantities, we have that
\begin{eqnarray}
&\Sigma^{ab}\!=\!-\frac{1}{2}\varepsilon^{abij}M_{ij}\\
&M^{ab}\!=\!\frac{1}{2}\varepsilon^{abij}\Sigma_{ij}
\end{eqnarray}
showing that only one of the two tensors is needed, and 
\begin{eqnarray}
&M_{ab}\Phi\!-\!\Sigma_{ab}\Theta\!=\!U^{j}S^{k}\varepsilon_{jkab}\label{A1}\\
&M_{ab}\Theta\!+\!\Sigma_{ab}\Phi\!=\!U_{[a}S_{b]}\label{A2}
\end{eqnarray}
together with
\begin{eqnarray}
&\frac{1}{2}M_{ab}M^{ab}\!=\!-\frac{1}{2}\Sigma_{ab}\Sigma^{ab}\!=\!\Phi^{2}\!-\!\Theta^{2}
\label{norm2}\\
&\frac{1}{2}M_{ab}\Sigma^{ab}\!=\!-2\Theta\Phi
\label{orthogonal2}\\
&U_{a}U^{a}\!=\!-S_{a}S^{a}\!=\!\Theta^{2}\!+\!\Phi^{2}\label{norm1}\\
&U_{a}S^{a}\!=\!0\label{orthogonal1}
\end{eqnarray}
called Fierz re-arrangement identities.

To classify spinorial fields, we first consider the case in which at least one of the two scalars is non-zero: in this case (\ref{norm1}) tells that $U^{a}$ is time-like, so that we can always perform up to three boosts in order to bring its spatial components to vanish; then it is always possible to use up to two rotations to bring the space part of $S^{a}$ aligned with the third axis; finally it is always possible to employ the last rotation to bring the spinor into the form
\begin{eqnarray}
&\!\!\psi\!=\!\phi e^{-\frac{i}{2}\beta\boldsymbol{\pi}}
\boldsymbol{S}\left(\!\begin{tabular}{c}
$1$\\
$0$\\
$1$\\
$0$
\end{tabular}\!\right)
\label{regular}
\end{eqnarray}
where $\boldsymbol{S}$ is a general spinor transformation. The complementary situation is given when both scalars are identically zero: in this case (\ref{norm2}, \ref{orthogonal2}) tell that if $M_{ab}$ is written in terms of $M_{0K}\!=\!E_{K}$ and $M_{IJ}\!=\!\varepsilon_{IJK}B^{K}$ then they are such that $E^{2}\!=\!B^{2}$ and $\vec{E}\!\cdot\!\vec{B}\!=\!0$ and employing the same reasoning used above we have that we can always boost and rotate so to bring $\vec{B}$ and $\vec{E}$ aligned respectively with the first and second axis; then, the spinor field is
\begin{eqnarray}
&\!\!\psi\!=\!e^{i\chi}
\boldsymbol{S}\left(\!\begin{tabular}{c}
$\cos{\frac{\theta}{2}}$\\
$0$\\
$0$\\
$\sin{\frac{\theta}{2}}$
\end{tabular}\!\right)
\label{singular1}
\end{eqnarray}
with $\boldsymbol{S}$ a general spin transformation. In either case the spinor field is said to be a spinor field in \emph{polar form} and henceforth we must specify whether it is the polar form of the regular spinor fields or the singular spinor fields.

In polar form for regular spinor fields
\begin{eqnarray}
&\Sigma^{ab}\!=\!2\phi^{2}(\cos{\beta}u^{[a}s^{b]}\!-\!\sin{\beta}u_{j}s_{k}\varepsilon^{jkab})\\
&M^{ab}\!=\!2\phi^{2}(\cos{\beta}u_{j}s_{k}\varepsilon^{jkab}\!+\!\sin{\beta}u^{[a}s^{b]})
\end{eqnarray}
showing that for regular spinors the tensors are written in terms of the vectors
\begin{eqnarray}
&S^{a}\!=\!2\phi^{2}s^{a}\\
&U^{a}\!=\!2\phi^{2}u^{a}
\end{eqnarray}
and the scalars
\begin{eqnarray}
&\Theta\!=\!2\phi^{2}\sin{\beta}\\
&\Phi\!=\!2\phi^{2}\cos{\beta}
\end{eqnarray}
in terms of a pseudo-scalar $\beta$ known as Yvon-Takabayashi angle and a scalar $\phi$ called module. The polar form of a singular spinor field shows that the tensors can not be written in terms of the other quantities, but we have 
\begin{eqnarray}
&S^{a}\!=\!-\cos{\theta}U^{a}
\end{eqnarray}
showing that $\theta$ is a function that describes the projection of the spin axial vector on the velocity vector, and thus that the spin axial vector is related to the Pauli-Lubanski axial vector while $\theta$ is linked to the helicity. For a detailed account on the classification in polar form see \cite{Fabbri:2016msm}.

From the metric, we define the symmetric connection as usual with $\Lambda^{\sigma}_{\alpha\nu}$ from which, with the tetrads, we define the spin connection $\Omega^{a}_{b\pi}\!=\!\xi^{\nu}_{b}\xi^{a}_{\sigma}(\Lambda^{\sigma}_{\nu\pi}\!-\!\xi^{\sigma}_{i}\partial_{\pi}\xi_{\nu}^{i})$ from which, with the gauge potential, we define the spinor connection
\begin{eqnarray}
&\boldsymbol{\Omega}_{\mu}
=\frac{1}{2}\Omega^{ab}_{\phantom{ab}\mu}\boldsymbol{\sigma}_{ab}
\!+\!iqA_{\mu}\boldsymbol{\mathbb{I}}\label{spinorialconnection}
\end{eqnarray}
in general. With it we can define
\begin{eqnarray}
&\boldsymbol{\nabla}_{\mu}\psi\!=\!\partial_{\mu}\psi
\!+\!\boldsymbol{\Omega}_{\mu}\psi\label{spincovder}
\end{eqnarray}
as the spinorial covariant derivative.

With spinor fields in polar form it is not difficult to see that $\boldsymbol{S}$ is generally given with the structure
\begin{eqnarray}
&\boldsymbol{S}\partial_{\mu}\boldsymbol{S}^{-1}\!=\!i\partial_{\mu}\sigma\mathbb{I}
\!+\!\frac{1}{2}\partial_{\mu}\theta_{ij}\boldsymbol{\sigma}^{ij}\label{parameters}
\end{eqnarray}
where $\sigma$ is a generic complex phase and $\theta_{ij}\!=\!-\theta_{ji}$ are the six parameters of the Lorentz group: then we can define
\begin{eqnarray}
&\partial_{\mu}\theta_{ij}\!-\!\Omega_{ij\mu}\!\equiv\!R_{ij\mu}\label{R}\\
&\partial_{\mu}\sigma\!-\!qA_{\mu}\!\equiv\!P_{\mu}\label{P}
\end{eqnarray}
which can be proven to be real tensors and therefore they are called \emph{tensorial connection} and \emph{gauge-invariant vector momentum}. With them, for regular spinor fields
\begin{eqnarray}
&\!\boldsymbol{\nabla}_{\mu}\psi\!=\!(\nabla_{\mu}\ln{\phi}\mathbb{I}
\!-\!\frac{i}{2}\nabla_{\mu}\beta\boldsymbol{\pi}
\!-\!iP_{\mu}\mathbb{I}\!-\!\frac{1}{2}R_{ij\mu}\boldsymbol{\sigma}^{ij})\psi
\label{decspinder}
\end{eqnarray}
in general. For singular spinor fields
\begin{eqnarray}
\nonumber
&\boldsymbol{\nabla}_{\mu}\psi\!=\!(-\frac{i}{2}\nabla_{\mu}\theta 
\boldsymbol{S}\boldsymbol{\gamma}^{2}\boldsymbol{S}^{-1}\boldsymbol{\pi}+\\
&+i\nabla_{\mu}\chi\!-\!iP_{\mu}\mathbb{I}\!-\!\frac{1}{2}R_{ij\mu}\boldsymbol{\sigma}^{ij})\psi
\end{eqnarray}
in which a strange $\boldsymbol{S}\boldsymbol{\gamma}^{2}\boldsymbol{S}^{-1}$ term appeared. In both cases
\begin{eqnarray}
&\nabla_{\mu}s_{i}\!=\!R_{ji\mu}s^{j}\label{ds}\\
&\nabla_{\mu}u_{i}\!=\!R_{ji\mu}u^{j}\label{du}
\end{eqnarray}
are valid as general geometric identities.

Thus far everything is developed in parallel for both of the two main Lounesto classes. If we want to specify onto the regular spinor fields, it is necessary to see in what way the polar form of the Dirac spinor is transmitted to the Dirac spinor field equations
\begin{eqnarray}
&i\boldsymbol{\gamma}^{\mu}\boldsymbol{\nabla}_{\mu}\psi
\!-\!XW_{\sigma}\boldsymbol{\gamma}^{\sigma}\boldsymbol{\pi}\psi\!-\!m\psi\!=\!0
\label{D}
\end{eqnarray}
where $W_{\nu}$ is the torsion axial vector.

Multiplying these field equations on the left by all Clifford matrices $\mathbb{I}$, $\boldsymbol{\pi}$, $\boldsymbol{\gamma}^{i}$, $\boldsymbol{\gamma}^{i}\boldsymbol{\pi}$, $\boldsymbol{\sigma}^{ij}$ and the adjoint spinor field, then splitting real and imaginary parts, we obtain
\begin{eqnarray}
&\frac{i}{2}(\overline{\psi}\boldsymbol{\gamma}^{\mu}\boldsymbol{\nabla}_{\mu}\psi
\!-\!\boldsymbol{\nabla}_{\mu}\overline{\psi}\boldsymbol{\gamma}^{\mu}\psi)
\!-\!XW_{\sigma}S^{\sigma}\!-\!m\Phi\!=\!0\\
&\nabla_{\mu}U^{\mu}\!=\!0
\end{eqnarray}
\begin{eqnarray}
&\frac{i}{2}(\overline{\psi}\boldsymbol{\gamma}^{\mu}\boldsymbol{\pi}\boldsymbol{\nabla}_{\mu}\psi
\!-\!\boldsymbol{\nabla}_{\mu}\overline{\psi}\boldsymbol{\gamma}^{\mu}\boldsymbol{\pi}\psi)
\!-\!XW_{\sigma}U^{\sigma}\!=\!0\\
&\nabla_{\mu}S^{\mu}\!-\!2m\Theta\!=\!0
\end{eqnarray}
\begin{eqnarray}
\nonumber
&\frac{i}{2}(\overline{\psi}\boldsymbol{\nabla}^{\alpha}\psi
\!-\!\boldsymbol{\nabla}^{\alpha}\overline{\psi}\psi)
\!-\!\frac{1}{2}\nabla_{\mu}M^{\mu\alpha}-\\
&-\frac{1}{2}XW_{\sigma}M_{\mu\nu}\varepsilon^{\mu\nu\sigma\alpha}\!-\!mU^{\alpha}\!=\!0
\label{vr}\\
\nonumber
&\nabla_{\alpha}\Phi
\!-\!2(\overline{\psi}\boldsymbol{\sigma}_{\mu\alpha}\!\boldsymbol{\nabla}^{\mu}\psi
\!-\!\!\boldsymbol{\nabla}^{\mu}\overline{\psi}\boldsymbol{\sigma}_{\mu\alpha}\psi)+\\
&+2X\Theta W_{\alpha}\!=\!0\label{vi}
\end{eqnarray}
\begin{eqnarray}
\nonumber
&\nabla_{\nu}\Theta\!-\!
2i(\overline{\psi}\boldsymbol{\sigma}_{\mu\nu}\boldsymbol{\pi}\boldsymbol{\nabla}^{\mu}\psi\!-\!
\boldsymbol{\nabla}^{\mu}\overline{\psi}\boldsymbol{\sigma}_{\mu\nu}\boldsymbol{\pi}\psi)-\\
&-2X\Phi W_{\nu}\!+\!2mS_{\nu}\!=\!0\label{ar}\\
\nonumber
&(\boldsymbol{\nabla}_{\alpha}\overline{\psi}\boldsymbol{\pi}\psi
\!-\!\overline{\psi}\boldsymbol{\pi}\boldsymbol{\nabla}_{\alpha}\psi)
\!-\!\frac{1}{2}\nabla^{\mu}M^{\rho\sigma}\varepsilon_{\rho\sigma\mu\alpha}+\\
&+2XW^{\mu}M_{\mu\alpha}\!=\!0\label{ai}
\end{eqnarray}
\begin{eqnarray}
\nonumber
&\nabla^{\mu}S^{\rho}\varepsilon_{\mu\rho\alpha\nu}
\!+\!i(\overline{\psi}\boldsymbol{\gamma}_{[\alpha}\!\boldsymbol{\nabla}_{\nu]}\psi
\!-\!\!\boldsymbol{\nabla}_{[\nu}\overline{\psi}\boldsymbol{\gamma}_{\alpha]}\psi)+\\
&+2XW_{[\alpha}S_{\nu]}\!=\!0\\
\nonumber
&\nabla^{[\alpha}U^{\nu]}\!+\!i\varepsilon^{\alpha\nu\mu\rho}
(\overline{\psi}\boldsymbol{\gamma}_{\rho}\boldsymbol{\pi}\!\boldsymbol{\nabla}_{\mu}\psi\!-\!\!
\boldsymbol{\nabla}_{\mu}\overline{\psi}\boldsymbol{\gamma}_{\rho}\boldsymbol{\pi}\psi)-\\
&-2XW_{\sigma}U_{\rho}\varepsilon^{\alpha\nu\sigma\rho}\!-\!2mM^{\alpha\nu}\!=\!0
\end{eqnarray}
which are known as Gordon-Madelung decompositions.

Finally, by substituting the polar form of the spinorial covariant derivative, after some manipulation and setting
\begin{eqnarray}
&\frac{1}{2}\varepsilon_{\mu\alpha\nu\iota}R^{\alpha\nu\iota}\!=\!B_{\mu}\\
&R_{\mu a}^{\phantom{\mu a}a}\!=\!R_{\mu}
\end{eqnarray}
one can isolate the pair of independent field equations
\begin{eqnarray}
&B_{\mu}\!-\!2P^{\iota}u_{[\iota}s_{\mu]}
\!-\!2XW_{\mu}\!+\!\nabla_{\mu}\beta\!+\!2s_{\mu}m\cos{\beta}\!=\!0\label{f1}\\
&R_{\mu}\!-\!2P^{\rho}u^{\nu}s^{\alpha}\varepsilon_{\mu\rho\nu\alpha}
\!+\!2s_{\mu}m\sin{\beta}\!+\!\nabla_{\mu}\ln{\phi^{2}}\!=\!0\label{f2}
\end{eqnarray}
which can be proven to imply (\ref{D}) and so demonstrating that (\ref{f1}, \ref{f2}) are together equivalent to the original Dirac spinor field equations. For more details on the derivation and general comments on the decomposition see \cite{Fabbri:2016laz}.

A general review on regular spinors is in \cite{Fabbri:2017pwp}.
\section{Flag Di-Pole Spinors}
So much as regular spinor fields, in their describing the Dirac spinor field, are important, the singular spinor field is also central, and this role is worth being addressed in terms of the polar decomposition. This was started in reference \cite{Fabbri:2017xyk}, but in this paper the analysis was essentially algebraic: here we plan to extend it to the dynamics.

The polar form of singular spinor fields
\begin{eqnarray}
&\!\!\psi\!=\!e^{i\chi}
\boldsymbol{S}\left(\!\begin{tabular}{c}
$\cos{\frac{\theta}{2}}$\\
$0$\\
$0$\\
$\sin{\frac{\theta}{2}}$
\end{tabular}\!\right)
\label{singular2}
\end{eqnarray}
contains an element that was not found for regular spinor fields, and that is that in the spinor field we cannot factor out the $\theta$ angle in the same way in which we factored out the Yvon-Takabayashi angle, although relationships like 
\begin{eqnarray}
&S^{a}\!=\!-\cos{\theta}U^{a}
\end{eqnarray}
show that the shifted function $\pi/2\!-\!\theta$ is a pseudo-scalar.

This fact is the reason why in 
\begin{eqnarray}
\nonumber
&\boldsymbol{\nabla}_{\mu}\psi\!=\!(-\frac{i}{2}\nabla_{\mu}\theta 
\boldsymbol{S}\boldsymbol{\gamma}^{2}\boldsymbol{S}^{-1}\boldsymbol{\pi}+\\
&+i\nabla_{\mu}\chi\!-\!iP_{\mu}\mathbb{I}\!-\!\frac{1}{2}R_{ij\mu}\boldsymbol{\sigma}^{ij})\psi
\label{decspindersing}
\end{eqnarray}
the strange $\boldsymbol{S}\boldsymbol{\gamma}^{2}\boldsymbol{S}^{-1}$ term appeared: because $\boldsymbol{S}$ and $\boldsymbol{\gamma}^{2}$ do not commute, there is no way to eliminate $\boldsymbol{S}$ in general circumstances. Therefore, by performing a spin transformation we will always be capable of changing the value of a term that should otherwise be invariant for such transformation. This term then violates Lorentz symmetry.

But there is more to it. As we have already mentioned, the singular spinors are further split in three sub-classes, one of which being the most general flag di-pole spinors, while the others are defined in terms of additional constraints like $S^{a}\!=\!0$ and containing the Majorana spinors and also $M^{ab}\!=\!0$ and containing the Weyl spinors. There can be no class with $U^{a}\!=\!0$ because in this case we would have the vanishing of the time component, which in turn would imply the vanishing of the whole spinor.

Flag di-pole spinor fields, in their generality, may have a $\theta$ angle that is itself a field, and therefore the strange-looking $\boldsymbol{S}\boldsymbol{\gamma}^{2}\boldsymbol{S}^{-1}$ term might always appear in the covariant derivative of the spinor field: because $\theta\!=\!0$ or $\theta\!=\!\pi$ are the values identifying a Weyl spinor while $\theta\!=\!\pm\pi/2$ are the values identifying a Majorana spinor, it could always be possible to have a Weyl spinor mutating into a Majorana spinor or viceversa. As intriguing as this situation may look, we admittedly never met anything alike.

The investigation of Flag di-pole spinor fields in polar form is obviously an open avenue of research. Nevertheless, for the moment we will focus on a possible mechanism that could seal off the two last Lounesto sub-classes.

This can be done by assuming that for some reason the $\theta$ angle were a constant, thus obtaining
\begin{eqnarray}
&\boldsymbol{\nabla}_{\mu}\psi\!=\!(i\nabla_{\mu}\chi\!-\!iP_{\mu}\mathbb{I}\!-\!\frac{1}{2}R_{ij\mu}\boldsymbol{\sigma}^{ij})\psi
\end{eqnarray}
which also no longer contains the frame-dependent factor, and this is spinorial covariant derivative we will use next.

Before that however, it is important to notice that the phase term has the same structure of the gauge-invariant momentum vector, and consequently it is with no loss of generality that we can write
\begin{eqnarray}
&\boldsymbol{\nabla}_{\mu}\psi\!=\!-(iP_{\mu}\mathbb{I}
\!+\!\frac{1}{2}R_{ij\mu}\boldsymbol{\sigma}^{ij})\psi
\label{aux}
\end{eqnarray}
since the phase can always be re-defined as to be absorbed into the gauge-invariant vector momentum itself. Hence, the spinorial covariant derivatives are written in terms of the gauge-invariant vector momentum and the tensorial connection alone. This fact looks bizarre because it seems to indicate that we have lost all dynamical characters.

To better understand this point, it is necessary to recall that when we perform the transformations leading to the polar form of spinor fields, these spinors are fields in the sense that they are generally point-dependent: the fact that it is fundamental to use local spin transformations $\boldsymbol{S}$ is precisely to ensure that the transformation could take place point-by-point. On the other hand, this means that the spinorial connection $\boldsymbol{\Omega}_{\mu}$ gets non-zero components.

These non-zero components are where the components of the original spinor field are transferred. In them it is where the information about the frame needed to perform the transfer to the polar form is stored, and when they are combined with the connection and potentials, themselves containing information about frames, we obtain that the frame dependence compensates and thus (\ref{R}, \ref{P}) can be proven to be tensors and gauge-invariant vectors \cite{Fabbri:2018crr}.

The tensorial connection and the gauge-invariant vector momentum are therefore objects that contain information about the space-time and gauge structures, not about the spinor field itself, whose true degrees of freedom cannot be transferred away. This is the reason why for regular spinors (\ref{decspinder}) contains the derivatives of module and Yvon-Takabayashi angle. But this is also the reason why for singular spinors the fact that (\ref{aux}) contains none of these elements appears to be a dynamical restriction.
\section{The Weyl Spinor}
Sealing off Majorana spinors from Weyl spinors allows us to study the two cases separately. We will begin with Weyl spinors, leaving Majorana spinors to a later section.

To study Weyl spinors, we start by recalling that they are single-handed (so either right-handed or left-handed) spinorial fields: this means that $M^{ab}\!=\!0$ and henceforth also $S^{a}\!=\!\pm U^{a}$ (with the plus or minus sign indicating a right-handed or a left-handed case). So Weyl spinors are cases of full helicity definition (either a totally parallel or a totally antiparallel) spinors. According to this we must expect them to be representing massless particles.

Inserting (\ref{aux}) into the Gordon-Madelung decompositions of the Dirac spinor field equations we find
\begin{eqnarray}
&R_{\mu}U^{\mu}\!=\!0\\
&(-B_{\mu}\!+\!2XW_{\mu}\!\pm\!2P_{\mu})U^{\mu}\!=\!0\\
&[(-B_{\mu}\!+\!2XW_{\mu}\!\pm\!2P_{\mu})\varepsilon^{\mu\rho\alpha\nu}
\!+\!g^{\rho[\alpha}R^{\nu]}]U_{\rho}\!=\!0
\end{eqnarray}
in terms of the vector current and where the masslessness constraint $m\!\equiv\!0$ emerges in a natural manner.

By writing these equations in the polar frame, the one where $\boldsymbol{S}\!=\!\mathbb{I}$ identically, it is possible to see that
\begin{eqnarray}
&U_{0}\!=\!U_{3}\!=\!1
\end{eqnarray}
and therefore we can expand the above equations into
\begin{eqnarray}
&R_{0}\!=\!R_{3}\\
&(-B\!+\!2XW\!\pm\!2P)_{0}\!=\!(-B\!+\!2XW\!\pm\!2P)_{3}\\
&(-B\!+\!2XW\!\pm\!2P)_{2}=-R_{1}\\
&(-B\!+\!2XW\!\pm\!2P)_{1}=R_{2}
\end{eqnarray}
as a constraint between components of the external fields thus showing no dynamical features, as expected.

We also remark that we have only $4$ equations compatibly with the fact that Weyl spinors have half components.
\section{Charge Flip}
The case of Majorana spinors is more intriguing because they are charge-conjugated fields, that is subject to the condition $\eta\boldsymbol{\gamma}^{2}\psi^{\ast}\!=\!\psi$ where $\eta$ is an irrelevant constant phase: in this case $S^{a}\!=\!0$ identically. Hence a Majorana spinor has neither electrodynamic coupling, because it is neutral, nor torsional coupling, since torsion would couple to the spin axial vector. This reduces the complexity.

In fact, because condition $\eta\boldsymbol{\gamma}^{2}\psi^{\ast}\!=\!\psi$ nullifies the possibility to perform any complex unitary transformation then no phase is admitted and we have only
\begin{eqnarray}
&\boldsymbol{\nabla}_{\mu}\psi\!=\!-\frac{1}{2}R_{ij\mu}\boldsymbol{\sigma}^{ij}\psi
\label{m}
\end{eqnarray}
as expression for the spinorial covariant derivative.

With it, the Gordon-Madelung decompositions are
\begin{eqnarray}
&(g_{\sigma[\pi}B_{\kappa]}\!-\!R^{\mu}\varepsilon_{\mu\sigma\pi\kappa})M^{\pi\kappa}\!=\!0\\
&\frac{1}{2}(B_{\mu}\varepsilon^{\mu\sigma\pi\kappa}
\!+\!g^{\sigma[\pi}R^{\kappa]})M_{\pi\kappa}\!-\!2mU^{\sigma}\!=\!0
\end{eqnarray}
in terms of the vector current and the tensor.

As above, in the polar frame we have
\begin{eqnarray}
&U_{0}\!=\!U_{3}\!=\!1\\
&M^{02}\!=\!M^{23}\!=\!1
\end{eqnarray}
and we can expand the above equations into
\begin{eqnarray}
&R_{0}\!=\!R_{3}\\
&B_{0}\!=\!B_{3}\\
&B_{2}=R_{1}\\
&-B_{1}=R_{2}\!-\!2m
\label{mass}
\end{eqnarray}
as a constraint between components of the external fields thus showing no dynamical features, again as expected.

As before we are down to only $4$ equations compatibly with the fact that a Majorana spinor has constraints.

It is also important to stress that a massless Majorana spinor and the special case of a Weyl spinor interacting only with the gravitational field are characterized by exactly the same structure of dynamical field equations.

Above, we have commented that Dirac types of spinor field equations for Majorana spinors fail to have massive solutions in the form of plane waves: this can be shown by taking the Dirac spinor field equation for the momenta
\begin{eqnarray}
&P_{\mu}\boldsymbol{\gamma}^{\mu}\psi\!-\!m\psi\!=\!0
\end{eqnarray}
under the Majorana constraint $\eta\boldsymbol{\gamma}^{2}\psi^{\ast}\!=\!\psi$ which gives
\begin{eqnarray}
&-\eta P_{\mu}\boldsymbol{\gamma}^{\mu}\boldsymbol{\gamma}^{2}\psi^{\ast}
\!-\!\eta m\boldsymbol{\gamma}^{2}\psi^{\ast}\!=\!0
\end{eqnarray}
and then 
\begin{eqnarray}
&P_{\mu}\boldsymbol{\gamma}^{\mu}\psi\!+\!m\psi\!=\!0
\end{eqnarray}
which implies $m\!=\!0$ identically. And as (\ref{mass}) shows, using only plane waves means $R_{ij\mu}\!=\!0$ and thus $m\!\equiv\!0$ holds.

As a matter of fact, the very condition of plane waves
\begin{eqnarray}
&i\boldsymbol{\nabla}_{\mu}\psi\!=\!P_{\mu}\psi
\end{eqnarray}
under the Majorana constraint becomes
\begin{eqnarray}
&i\boldsymbol{\nabla}_{\mu}\psi\!=\!-P_{\mu}\psi
\end{eqnarray}
giving $P_{\mu}\!=\!0$ and showing that Majorana spinors can not be written with the structure of plane-wave solutions.

This points toward the fact that if one wants to develop a theory of Majorana spinors then the dynamics can not be given by a Dirac type of spinor field equation.

Different ways out are possible: the most common one, that is the solution adopted in quantum field theory, has at its basis the idea of taking Majorana spinors with the structure of Grassmann-valued fields. However, this consists in changing the algebraic structure of the spinor, not the differential structure of the propagating spinor itself.

Another solution would then be to change the structure of the dynamics by changing the field equations. Because the Dirac equation is the most general we can have at the first-order derivative then the solution may only be to go at the second-order derivative in the field equations.

In the next section we will investigate such a case.
\section{ELKO}
When considering second-order derivative field equations, we are essentially taking Klein-Gordon field equations. Therefore this field is a spinor with a unitary mass dimension \cite{Ahluwalia:2004sz,Ahluwalia:2004ab,Ahluwalia:2016rwl,Ahluwalia:2016jwz}. Such a type of scalar-like Majorana spinor is known as ELKO. As Majorana spinors they are
\begin{eqnarray}
&\lambda^{S}\!=\!\left(\!\begin{tabular}{c}
$\boldsymbol{\sigma}^{2}R^{\ast}$\\
$R$
\end{tabular}\!\right)\ \ \ \ \mathrm{or}\ \ \ \ 
&\lambda^{A}\!=\!\left(\!\begin{tabular}{c}
$-\boldsymbol{\sigma}^{2}R^{\ast}$\\
$R$
\end{tabular}\!\right)
\end{eqnarray}
as Self-conjugate or Antiself-conjugate ELKO. And being scalar-like fields they verify the field equations given by
\begin{eqnarray}
&\boldsymbol{\nabla}^{2}\lambda\!+\!m^{2}\lambda\!=\!0
\label{e}
\end{eqnarray}
which for now are taken in free case.

The polar form of ELKO is of course
\begin{eqnarray}
&\lambda\!=\!\boldsymbol{S}\left(\!\begin{tabular}{c}
$1$\\
$0$\\
$0$\\
$1$
\end{tabular}\!\right)
\end{eqnarray}
then giving
\begin{eqnarray}
&\boldsymbol{\nabla}_{\mu}\lambda\!=\!-\frac{1}{2}R_{ij\mu}\boldsymbol{\sigma}^{ij}\lambda
\end{eqnarray}
in terms of the tensorial connection.

By plugging it into (\ref{e}) we get the polar form
\begin{eqnarray}
\nonumber
&[\frac{1}{2}\nabla_{\mu}R^{ij\mu}\boldsymbol{\sigma}_{ij}
\!-\!\frac{i}{16}R_{ij\nu}R_{pq}^{\phantom{pq}\nu}\varepsilon^{ijpq}\boldsymbol{\pi}+\\
&+(\frac{1}{8}R_{ij\nu}R^{ij\nu}\!-\!m^{2})\mathbb{I}]\lambda\!=\!0
\end{eqnarray}
which contains no derivatives of the spinor, and therefore the Madelung-Gordon decomposition would result into a number of constraints on the tensorial connection.

By taking second-order field equations we might raise the constraints on the tensorial connection from algebraic to differential, but we still have no dynamical feature.

This is obvious, since the polar form of ELKO has no degree of freedom left. As a consequence, we may wonder how this problem can be solved, and the solution comes from a detail that for ELKO must be considered although so far in the treatment we have not considered it yet.

A problem that must be faced in the case of ELKO is that, as said above, they belong to the singular Lounesto class, so both their scalars vanish: hoping to define some bi-linear scalars that are not trivial, another adjunction, or dual, must be introduced. This dual can be built taking 
$\widetilde{\lambda}\!=\!\lambda^{\dagger}\boldsymbol{\gamma}^{0}\boldsymbol{A}$ for some 
$\boldsymbol{A}$ to be found. We have noticed that this matrix is bound to be either $\boldsymbol{\pi}$ or $\mathbb{I}$ and so the new dual can be constructed only by relaxing some of the implicitly imposed conditions. One very subtle condition, for instance, is that we are assuming that a spinor dual is written in terms of the initial spinor alone. However, this may not always be the case, as shown by Ahluwalia.

The idea is to take advantage of the fact that ELKO, apart from self-conjugate or antiself-conjugate, also have another type of duplicity given by the fact that $R$ can be either in spin-up or spin-down configuration: indicating with a plus the spin-up and with a minus the spin-down we have that for each ELKO we can construct
\begin{eqnarray}
&\lambda^{S}_{\pm}\!=\!\left(\!\begin{tabular}{c}
$\boldsymbol{\sigma}^{2}R^{\ast}_{\pm}$\\
$R_{\pm}$
\end{tabular}\!\right)\ \ \ \ \mathrm{or}\ \ \ \ 
&\lambda^{A}_{\pm}\!=\!\left(\!\begin{tabular}{c}
$-\boldsymbol{\sigma}^{2}R^{\ast}_{\mp}$\\
$R_{\mp}$
\end{tabular}\!\right)
\end{eqnarray}
or alternatively
\begin{eqnarray}
&\lambda^{S}_{+}\!=\!\left(\!\begin{tabular}{c}
$0$\\
$iBe^{-i\beta}$\\
$Be^{i\beta}$\\
$0$
\end{tabular}\!\right)\ \ \ \ \mathrm{and} \ \ 
&\lambda^{S}_{-}\!=\!\left(\!\begin{tabular}{c}
$-iBe^{-i\beta}$\\
$0$\\
$0$\\
$Be^{i\beta}$
\end{tabular}\!\right)
\end{eqnarray}
\begin{eqnarray}
&\lambda^{A}_{+}\!=\!\left(\!\begin{tabular}{c}
$iBe^{-i\beta}$\\
$0$\\
$0$\\
$Be^{i\beta}$
\end{tabular}\!\right)\ \ \ \ \mathrm{and} \ \ 
&\lambda^{A}_{-}\!=\!\left(\!\begin{tabular}{c}
$0$\\
$-iBe^{-i\beta}$\\
$Be^{i\beta}$\\
$0$
\end{tabular}\!\right)
\end{eqnarray}
for a total amount of four independent spinors. The dual is now defined in terms of the usual procedure but mixing different helicity eigen-states according to 
\begin{eqnarray}
&\widetilde{\lambda}^{S,A}_{\pm}\!=\!\mp i\left(\lambda^{S,A}_{\mp}\right)^{\dagger} \boldsymbol{\gamma}^{0}
\end{eqnarray}
so that
\begin{eqnarray}
&\widetilde{\lambda}^{S}_{+}\!=\!\left(\!\begin{tabular}{cccc}
$0$\ \ & $-iBe^{-i\beta}$\ \  & $Be^{i\beta}$\ \  & $0$
\end{tabular}\!\right)\\
&\widetilde{\lambda}^{S}_{-}\!=\!\left(\!\begin{tabular}{cccc}
$iBe^{-i\beta}$\ \  & $0$\ \  & $0$\ \  & $Be^{i\beta}$ 
\end{tabular}\!\right)
\end{eqnarray}
\begin{eqnarray} 
&\widetilde{\lambda}^{A}_{+}\!=\!\left(\!\begin{tabular}{cccc}
$-iBe^{-i\beta}$\ \  & $0$\ \   & $0$\ \  & $Be^{i\beta}$
\end{tabular}\!\right)\\
&\widetilde{\lambda}^{A}_{-}\!=\!\left(\!\begin{tabular}{cccc}
$0$\ \  & $iBe^{-i\beta}$\ \   & $Be^{i\beta}$\ \  & $0$
\end{tabular}\!\right)
\end{eqnarray}
as it has been extensively discussed in references \cite{Ahluwalia:2016rwl,Ahluwalia:2016jwz}.

Notice that in this way the scalars are given by
\begin{eqnarray}
&\widetilde{\lambda}^{S}_{\pm}\lambda^{S}_{\pm}
\!=\!\widetilde{\lambda}^{A}_{\pm}\lambda^{A}_{\pm}
\!=\!2B^{2}\cos{(2\beta)}\label{s}\\
&i\widetilde{\lambda}^{S}_{\pm}\boldsymbol{\pi}\lambda^{S}_{\pm}
\!=\!i\widetilde{\lambda}^{A}_{\pm}\boldsymbol{\pi}\lambda^{A}_{\pm}
\!=\!-2B^{2}\sin{(2\beta)}\label{p}
\end{eqnarray}
which are in fact non-null, or not necessarily at least.

The definition of a new dual poses however some problem about the Lounesto classification, because this classification was based on the standard dual. The standard dual could of course still be used even if the new dual is defined, and so the Lounesto classification works just the same. However, to dissipate all doubts, we will show the polar decomposition for ELKO by using only the transformation laws given by the local spin transformations.

The local spin transformations are very well known and they do not change for ELKO: for instance
\begin{eqnarray}
\boldsymbol{S}_{R3}\!=\!\left(\begin{array}{cccc}
\!e^{i\zeta}\!&\!0\!&\!0\!&\!0\!\\ 
\!0\!&\!e^{-i\zeta}\!&\!0\!&\!0\!\\ 
\!0\!&\!0\!&\!e^{i\zeta}\!&\!0\!\\ 
\!0\!&\!0\!&\!0\!&\!e^{-i\zeta}\!\\
\end{array}\right)
\end{eqnarray}
for the rotation around the third axis and
\begin{eqnarray}
&\boldsymbol{S}_{B3}\!=\!\left(\begin{array}{cccc}
\!e^{-\eta}\!&\!0\!&\!0\!&\!0\!\\ 
\!0\!&\!e^{\eta}\!&\!0\!&\!0\!\\ 
\!0\!&\!0\!&\!e^{\eta}\!&\!0\!\\ 
\!0\!&\!0\!&\!0\!&\!e^{-\eta}\!\\
\end{array}\right)
\end{eqnarray}
for the boost along the third axis. For the ELKO we can take any of the four forms, so we pick for instance
\begin{eqnarray}
&\lambda^{S}_{-}\!=\!\left(\!\begin{tabular}{c}
$-iBe^{-i\beta}$\\
$0$\\
$0$\\
$Be^{i\beta}$
\end{tabular}\!\right)
\end{eqnarray}
knowing that for all others the results would not change.

By applying on $\lambda^{S}_{-}$ the rotation around the third axis with $\zeta\!=\!\beta\!+\!\pi/4$ and then the boost along the same axis with $\eta\!=\!\ln{B}$ we see that we can always write \begin{eqnarray}
&\lambda^{S}_{-}\!=\!e^{-i\pi/4}\boldsymbol{S}_{R3}^{-1}
\boldsymbol{S}_{B3}^{-1}\left(\!\begin{tabular}{c}
$1$\\
$0$\\
$0$\\
$1$
\end{tabular}\!\right)
\end{eqnarray}
which is a special case of the polar forms (\ref{singular1}) or (\ref{singular2}).

This result does not depend on which ELKO is taken and so we can always write ELKO in such a form, but one crucial point is that different ELKO would have different polar forms, so that we must write in general that
\begin{eqnarray}
&\lambda^{S}_{+}\!=\!\boldsymbol{S}^{S}_{+}\left(\!\begin{tabular}{c}
$0$\\
$1$\\
$1$\\
$0$
\end{tabular}\!\right)\ \ \ \ \lambda^{S}_{-}\!=\!\boldsymbol{S}^{S}_{-}\left(\!\begin{tabular}{c}
$1$\\
$0$\\
$0$\\
$1$
\end{tabular}\!\right)\\
&\lambda^{A}_{+}\!=\!\boldsymbol{S}^{A}_{+}\left(\!\begin{tabular}{c}
$1$\\
$0$\\
$0$\\
$1$
\end{tabular}\!\right)\ \ \ \ \lambda^{A}_{-}\!=\!\boldsymbol{S}^{A}_{-}\left(\!\begin{tabular}{c}
$0$\\
$1$\\
$1$\\
$0$
\end{tabular}\!\right)
\end{eqnarray}
where $\boldsymbol{S}^{S,A}_{\pm}$ is a local spin transformation, which depends on the particular ELKO to which it is correlated.

Therefore ELKO can have all their components transferred through a local spin transformation into the frame and as a consequence into the tensorial connection, but correspondingly we have that all different ELKO will be given in terms of different tensorial connections.

This point is of paramount importance because, while on the one hand, this is needed to ensure (\ref{s}, \ref{p}) to have scalar transformations, on the other hand, it also means that we can never write a polar form for all ELKO at the same time: the transformation that polarizes an ELKO is also the transformation that depolarizes the other ELKO helicity. Therefore we must conclude that ELKO can not admit a unique polar decomposition for all helicity states and consequently ELKO can have no polar form.

This ensures that ELKO will always have some degree of freedom, whether in one or the other helicity state.

On the other hand, this also means that there can be no polar form transmitted onto the ELKO field equations.

Converting a theory in polar form has the advantage that, by transferring as many components as possible into the tensorial connection, only the degrees of freedom will be left in the field, and that only real tensorial quantities appear, which means that the treatment is at its clearest.

So the question that naturally arises is whether it exists at least \emph{some} type of polar decomposition that can be successfully employed in the case of the ELKO.

We leave this question to a following work.
\section{Conclusion}
We have considered the polar form of the spinor fields studying their structure in different Lounesto classes.

We have done the general study for all Lounesto classes in parallel, exemplifying the treatment of regular spinor fields by means of the well known Dirac spinor fields.

For singular spinor fields, we gave general comments about the flag di-pole spinor fields, and in particular we presented the relationship $S^{a}\!=\!-\cos{\theta}U^{a}$ in terms of the $\theta$ parametric field, which is a parameter since it defines further sub-classes of flag di-pole, where Weyl and Majorana spinors are, but it is also a field since it varies, so leading to the fact that Weyl could mutate into Majorana spinors: this occurrence has never been seen in nature and therefore it makes sense to get rid of such a possibility in the further investigation of Lounesto classes.

By requiring $\theta$ constant, sealing off the two sub-classes, we were directed toward an independent study of these last two classes, consisting of Weyl and Majorana spinors: we have discussed how in these instances, all components of the spinors are transferred by the polar form into the frame, therefore into the tensorial connection. This looks strange because it indicates that no degree of freedom is left, neither for Weyl nor for Majorana spinors alike, and this circumstance seems unacceptable when developing a dynamics for this type of propagating spinor fields.

Weyl spinors were studied by finding the polar form of their field equations, which indeed had no derivatives for the degrees of freedom, but were merely constraints upon the tensorial connection itself. And the same was found for Majorana spinors. Therefore, Weyl and Majorana are dynamically very similar spinor fields indeed.

As Majorana spinors cannot have massive plane-wave solutions for Dirac-like spinor field equations, we went on to conclude that one might have to use a Klein-Gordon field equation: the Majorana spinor verifying scalar-like field equations is called ELKO, and we studied ELKO in the perspective of the polar decomposition. It was found that the polar form of field equations lacked derivatives of degrees of freedom, consisting in a number of constraints on the tensorial connection. By exploiting ELKO double helicity we saw that a new type of dual could be defined mixing ELKO helicities, in terms of which we have seen that ELKO can in fact maintain real degrees of freedom in polar form, but we have also seen that for ELKO no unique polar decomposition can in fact be constructed.

So what can we make out of this? Apart from the regular spinor fields, the situation for singular spinor fields seems rather cumbersome: flag di-pole spinor fields show features that do not occur in nature, unless some targeted restriction is imposed: but even in this case, whether for Weyl or for Majorana spinors, there is the unsettling circumstance that all of their components can be transferred by the polar transformation into the frame, leaving them with no degree of freedom but only with some tensorial connection, and thus non-dynamical. Like Majorana also ELKO have this strange non-dynamical character: a way out is to define the ELKO dual, because when an ELKO is written in polar form the dual ELKO cannot be written in the same polar form. But this also means ELKO have no unique polar form for all helicities, and thus that ELKO field equations cannot be given in polar form.

However, writing equations in polar form \emph{is} a considerable advantage, because by transferring as many components as possible into the tensorial connection, we keep only the degrees of freedom of the field in the field itself, and the treatment results to be clearest. So is it possible to have at least \emph{some} polar decomposition for ELKO?

And if yes, what will the polar field equations be?

\end{document}